\documentclass{article}
\usepackage{graphicx}
\usepackage[margin=3cm, a4paper]{geometry}
\usepackage{cite}
\usepackage{amsmath,amssymb,amsfonts}
\usepackage{siunitx}
\usepackage{graphicx}
\usepackage{subcaption}
\usepackage{overpic}
\usepackage{textcomp}
\usepackage{algorithm}
\usepackage{algpseudocode}
\usepackage{url}
\usepackage{booktabs}
\usepackage{colortbl}
\usepackage{makecell}
\usepackage{wasysym}
\usepackage{gensymb}
\usepackage{authblk}

\begin{document}

\title{Image Reconstruction in Cone Beam Computed Tomography Using Controlled Gradient Sparsity}

\author[1]{Alexander Meaney\footnote{Corresponding author. Email: alexander.meaney@helsinki.fi.}}
\author[2]{Mikael A. K. Brix}
\author[2]{Miika T. Nieminen}
\author[1]{\\Samuli Siltanen}
\affil[1]{{\small Department of Mathematics and Statistics, University of Helsinki, Helsinki, Finland}}
\affil[2]{{\small Research Unit of Health Sciences and Techonology, University of Oulu, and the Department of Diagnostics, Oulu University Hospital, Oulu, Finland}}

\date{}

\maketitle

\begin{abstract}
Total variation (TV) regularization is a popular reconstruction method for ill-posed imaging problems, and particularly useful for applications with piecewise constant targets. However, using TV for medical cone-beam computed X-ray tomography (CBCT) has been limited so far, mainly due to heavy computational loads at clinically relevant 3D resolutions and the difficulty in choosing the regularization parameter. Here an efficient minimization algorithm is presented, combined with a dynamic parameter adjustment based on control theory. The result is a fully automatic 3D reconstruction method running in clinically acceptable time. The input on top of projection data and system geometry is desired degree of sparsity of the reconstruction. This can be determined from an atlas of CT scans, or alternatively used as an easily adjustable parameter with straightforward interpretation.
\end{abstract}

\section{Introduction}
\label{sec:introduction}

Cone-beam computed tomography (CBCT) has become a standard imaging tool at medium and large dental clinics, and it is increasingly used in other radiological applications as well \cite{fahrig_et_al_2021__jmedimaging, kaasalainen_et_al_2021__physmed}. The FDK algorithm \cite{feldkamp_et_al_1984__joptsocamera}, introduced in 1984, remains the standard image reconstruction technique in CBCT. However, while FDK is fast and well understood, there are many cases where it is not the optimal solution, such as low-dose imaging, sparse-view or limited-angle imaging, out-of-plane distortion correction, 4D imaging, metal artifact reduction, and soft-tissue contrast enhancement. Consequently, in recent years there has been increasing interest in approaches based on machine learning and model-based iterative reconstruction (MBIR) \cite{gardner_et_al_2019__advradiatoncol, 
ravishankar_et_al_2020__procieee,
matenine_et_al_2020__neuroradiology, 
wu_et_al_2020__medphys, 
lagerwerf_et_al_2020__jimaging,
lu_et_al_2021__physmedbiol,
moriakov_et_al_2023__medphys}. 

MBIR techniques are often based on a variational approach where \emph{a priori} information on the desired solution is included in the reconstruction problem via a regularization function, and a regularization parameter $\alpha>0$ specifies the balance between data discrepancy and regularization. A commonly used regularization strategy is total variation (TV), which penalizes the $\ell^1$ norm of the image gradient.

TV regularization was first introduced in 1992 for image denoising \cite{rudin_et_al_1992__physicad}. Early applications in X-ray tomography were reported in \cite{delaney_bresler_1998__ieeetransimageprocess, persson_et_al_2001__physmedbiol}, and first results with measured data were presented in \cite{kolehmainen_et_al_2003__physmedbiol}. Several more studies followed \cite{kolehmainen_et_al_2006__ieeetransmedimaging, 
sidky_et_al_2006__nssmic, 
sidky_et_al_2006__jxrayscitechnol, 
liao_sapiro_2008__isbi, 
herman_davidi_2008__inverseprobl, 
tang_et_al_2009__physmedbiol, 
duan_et_al_2009__ieeetransnuclsci, 
bian_et_al_2010__tsinghuascitechnol, 
bian_et_al_2010__physmedbiol, 
tian_et_al_2011__physmedbiol,
jensen_et_al_2012__bitnumermath,
hamalainen_et_al_2014__intjtomogrsimul,
jorgensen_sidky_2015__philostransrsoca}, 
and TV-based solutions have become an often employed, albeit unstandardized, approach to tomographic reconstruction 
\cite{yang_et_al_2017__physmedbiol,
chen_et_al_2018__physmedbiol,
tseng_et_al_2020__physmed,
tseng_et_al_2022__physmedbiol,
sabate_landman_2023__physmedbiol,
zhang_et_al_2023__medphys}.
However, there remains the problem of choosing the optimal value for  $\alpha$. Several methods have been proposed for automatic choice 
\cite{clason_et_al_2010__siamjscicomput, 
dong_et_al_2011__jmathimagingvis, 
wen_chan_2012__ieeetransimageprocess, 
kolehmainen_et_al_2012__inverseprobl, 
frick_et_al_2012__electronjstat, 
kindermann_et_al_2014__jinverseillposedprobl, 
chen_et_al_2014__numeralgor, 
toma_et_al_2015__inverseproblimaging, 
niinimaki_et_al_2016__siamjimagingsci}, but they are impractical for CBCT as they require computing many reconstructions and retroactively choosing the optimal parameter, or are otherwise complicated to implement. Furthermore, there is little to no interpretability in the actual numerical value of $\alpha$.

We present an automatic TV parameter choice method based on adjusting the value of $\alpha$ adaptively during iteration so that the gradient magnitude sparsity of the reconstruction matches a predefined value. 
Choice of the regularization parameter based on an \emph{a priori} sparsity level was discussed in \cite{kolehmainen_et_al_2012__inverseprobl, hamalainen_et_al_2013__siamjscicomput}, and adaptive regularization was introduced in \cite{purisha_et_al_2018__measscitechnol} for wavelet-based sparsity reconstruction, where the percentage of (essentially) nonzero wavelet transform coefficients is known. In this work, we assume the percentage of voxels where the discrete gradient is (essentially) nonzero to be known. Such knowledge can be estimated for example from an atlas of CT reconstructions of relevant types of anatomy. We implement image reconstruction with an efficient primal-dual fixed point minimization method \cite{chen_et_al_2016__jfixpointtheorya}. The resulting algorithm is efficient enough to yield reconstructions at practical resolutions within practically feasible times. Furthermore, defining the regularization in terms of gradient sparsity greatly increases interpretability of the regularization strength.

Our work is based on two assumptions: 
\begin{enumerate}
    \item X-ray attenuation distribution within human tissues are approximately piecewise constant, and
    \item gradient sparsity of a certain type of anatomy (dentomaxillofacial, skull and brain, limb, \dots) is the same from one patient to another. 
\end{enumerate}
Our experiments with computational phantom data suggest that these assumptions might hold to a reasonable accuracy in clinical practice. Clinical CBCT scanners often operate with only a small set of options for reconstruction grid size and voxel resolution, which can aid in limiting the choice of sparsity for different imaging protocols.

\section{Methods}

\subsection{CBCT Imaging Model}

\noindent
The X-ray detector in a CBCT scanner is a 2D panel where the detector pixels are arranged into a $P_1 \times P_2$ array. Each pixel is labeled with an index $p \in \{1, \ldots, P \}$ where $P = P_1P_2$. The measurement at pixel $p$ at projection angle $\theta$ is defined as 
\begin{equation}
    \label{eq:pixel_measurement}
    m_{p, \theta} = -\ln\frac{I_{p, \theta}}{I_{p, 0}},
\end{equation}
where $I_{p, \theta}$ is the intensity measured by the pixel, and $I_{p, 0}$ is the intensity measured when no object is placed between the X-ray source and the detector. From Lambert-Beer's law it follows that 
\begin{equation}
    \label{eq:lambert_beer}
    -\ln\frac{I_{p, \theta}}{I_{p, 0}} = \int_\mathrm{ray\:path} \mathcal{F}(x, y, z) ds,
\end{equation}
where $\mathcal{F}(x, y, z)$ is the spatial distribution of attenuation coefficients in the scanned object \cite{kak_slaney_1988, buzug_2008}. The total number of data points collected in a single CBCT scan is $M = P\Theta$, where $\Theta$ is the number of projection angles. The data points are labeled $d \in \{1, \ldots, M \}$ and they are collected into the measurement data vector $m \in \mathbb{R}^M$.

We note that $\mathcal{F}(x, y, z)$ also depends on photon energy, and (\ref{eq:lambert_beer}) is strictly true for monochromatic X-ray sources only, whereas most clinical scanner use polychromatic radiation. However, most mathematical models for tomographic reconstruction use this model, and issues arising from polychromatic radiation are ignored or compensated for in other ways \cite{kak_slaney_1988, park_et_al_2015__philostransrsoca}.

$\mathcal{F}(x, y, z)$ is discretized as the $N_1 \times N_2 \times N_3$ voxel grid $f$, where the voxels are labeled as $f_{i,j,k}$ with $i \in \{1, \ldots, N_1\}$, $j \in \{1, \ldots, N_2\}$, and $k \in \{1, \ldots, N_3\}$. $f$ can also be flattened into a vector in $\mathbb{R}^N$ with $N = N_1 N_2 N_3$, as long as the location of each element is tracked when shifting between the 3D voxel and vector representations. In the vector representation, the elements are labeled $f_l$ with $l \in \{1, \ldots, N\}$. This work uses $f$ to refer to both the voxel and the vector representations, and it is assumed to be implicitly clear from the context which one applies.

The computational imaging model is obtained by discretizing the integral in (\ref{eq:lambert_beer}) as weighted sums across $f$ for each $m_d$ and collecting these as the rows of the matrix $A \in \mathbb{R}^{M \times N}$. $A$ is called the forward operator. This leads to the linear measurement model
\begin{equation}
    \label{eq:linear_measurement_model}
    Af = m + \varepsilon,
\end{equation}
where $\varepsilon \in \mathbb{R}^M$ represents noise and modeling error.

Eq. (\ref{eq:linear_measurement_model}) is an ill-posed problem, and iterative solvers that minimize $\|Af-m\|_2$ exhibit semi-convergence: the iterations initially approach a reasonable solution, but eventually converge towards meaningless noise \cite{natterer_1986}. The ill-posed problem is stabilized using a regularization function $R(f)$, which add \emph{a priori} knowledge of $f$. This leads to the regularized variational reconstruction problem
\begin{equation}
    \label{eq:regularized_ct_model}
        \min_{f\in\mathbb{R}^N} \frac{1}{2}\|Af-m\|_2^2 + \alpha R(f),
\end{equation}
where the regularization parameter $\alpha$ controls the strength of the regularization.

In most practical CBCT applications (\ref{eq:regularized_ct_model}) becomes a very large-scale optimization problem. For example, for a reconstruction volume of size $512 \times 512 \times 512$ voxels, $f$ contains approximately 134 million unknown values.

\subsection{Total Variation Regularization}
\label{sec:tv}

In total variation, the regularization function in (\ref{eq:regularized_ct_model}) is set as
\begin{equation}
    R(f) = \int_\Omega |\nabla f| \, dx \, dy \, dz,
\end{equation}
where $\Omega$ is the image domain \cite{rudin_et_al_1992__physicad, getreuer_2012__ipol}. In order to numerically solve the minimization problem, the gradient operator $\nabla$ needs to be discretized, which was done using finite forward differences. For computational purposes, it is useful to maintain a constant image size. We therefore define the numerical derivatives
\begin{align}
    \label{eq:dx}
        \partial_x f_{i, j, k} &=
        \begin{cases}
            f_{i, j + 1, k} - f_{i, j, k}, & j \in \{1, \ldots, N_2-1\}\\
            0, & j = N_2,
        \end{cases}        
        \\
    \label{eq:dy}
        \partial_y f_{i, j, k} &=
        \begin{cases}
            f_{i + 1, j, k} - f_{i, j, k}, & i \in \{1, \ldots, N_1-1\}\\
            0, & i = N_1,
        \end{cases}     
        \\
    \label{eq:dz}
        \partial_z f_{i, j, k} &=
        \begin{cases}
            f_{i, j, k+1} - f_{i, j, k}, & k \in \{1, \ldots, N_3-1\}\\
            0, & k = N_3,
        \end{cases}    
\end{align}
where the last voxels along each dimension are filled with zero-valued dummy variables \cite{condat_2017__siamjimagingsci}. The gradient at $f_{i,j,k}$ is
\begin{equation}
    \nabla f_{i, j, k} = 
    \begin{bmatrix} \partial_x f_{i, j, k} \\ \partial_y f_{i, j, k} \\ \partial_z f_{i, j, k}\end{bmatrix} 
\end{equation}
and the corresponding gradient magnitude is 
\begin{equation}
    \label{eq:gradient_magnitude}
    \|\nabla f_{i, j, k}\|_2 = \sqrt{(\partial_x f_{i, j, k})^2 + (\partial_y f_{i, j, k})^2 + (\partial_z f_{i, j, k})^2}.
\end{equation}
The TV regularization term now becomes
\begin{equation}
    \label{eq:discrete_tv_term}
    R(f) = \sum_{i=1}^{N_1} \sum_{j=1}^{N_2} \sum_{k=1}^{N_3} \|\nabla f_{i, j, k}\|_2.
\end{equation}
We use the unconventional notation $\|\nabla f\|_2 \in \mathbb{R}^N$ to refer to the vector containing all the gradient magnitude values of $f$. Noting that (\ref{eq:discrete_tv_term}) is the $\ell^1$ norm of $\|\nabla f\|_2$, the regularization function can be written using the mixed $\ell^{2,1}$ norm \cite{gramfort_et_al_2012__physmedbiol}:
\begin{equation}
    \label{eq:mixed_gradient_norm}
    R(f) = \|\nabla f\|_{2, 1}.
\end{equation}

Let $D_x, D_y, D_z \in \mathbb{R}^{N \times N}$ be the matrices that implement the linear maps defined (\ref{eq:dx}), (\ref{eq:dy}), and (\ref{eq:dz}), respectively. We now define the discrete gradient matrix $D \in \mathbb{R}^{3N \times N}$ as
\begin{equation}
    D = 
    \begin{bmatrix}
        D_x \\ D_y \\ D_z
    \end{bmatrix}.
\end{equation}
For the matrix-vector product $Df$, entries $\{1\ldots \,N\}$ encode the discrete $x$-derivative of $f$, entries $\{N + 1\ldots \,2N\}$ encode the discrete $y$-derivative of $f$, and entries $\{2N + 1\ldots \,3N\}$ encode the discrete $z$-derivative of $f$. Eq. (\ref{eq:mixed_gradient_norm}) can now be expressed as:
\begin{equation}
    R(f) = \| Df \|_{2, 1},
\end{equation}
provided that the computations keep track of the entries used to compute each norm \cite{chen_et_al_2013__inverseprobl}. 

An iterative solver for (\ref{eq:regularized_ct_model}) will involve numerous forward projections (multiplying with $A$) and backprojections (multiplying with $A^T$). In order to preserve the total intensity of the grayscale values of $f$ in such an operation, the linear operator $A$ should be replaced with the normalized forward model $\widetilde{A} = A/\|A\|_2$ and the measurement data with the normalized data $\widetilde{m} = m/\|A\|_2$ in the solver. Otherwise, the reconstructed voxel values will not correspond on average with those obtained from an FDK reconstruction, and at worst they can spiral toward infinity or toward zero. We emphasize that this normalization requirement only applies to implementing the iterative solver, not to the initial measurement model described in (\ref{eq:pixel_measurement})-(\ref{eq:linear_measurement_model}).

X-rays are only attenuated, never amplified, while passing through a me\-di\-um. Therefore the values of $f$ can be constrained to the nonnegative orthant $\mathbb{R}_+^N$. The final form of the computational reconstruction problem becomes
\begin{equation}
    \label{eq:regularized_ct_tv_model}
    \min_{f \in \mathbb{R}_+^N} \frac{1}{2}\|\widetilde{A}f - \widetilde{m}\|_2^2  + \alpha \|D f\|_{2, 1}.
\end{equation}

The primal-dual fixed point (PDFP) algorithm proposed by Chen, Huang, and Zhang \cite{chen_et_al_2016__jfixpointtheorya} is well suited to solving the large-scale optimization problem in (\ref{eq:regularized_ct_tv_model}). The PDFP algorithm solves minimization problems of the form 
\begin{equation}
    \label{eq:pdfp}
    \min_{f \in \mathbb{R}^N} L_1(f) + (L_2 \circ B)(f) + L_3(f),
\end{equation}
where $L_1$, $L_2$, and $L_3$ are three proper lower semi-continuous convex functions, $L_1$ is differentiable on $\mathbb{R}^N$ with a $1/\beta$-Lipschitz continuous gradient for some $\beta \in (0, \infty]$, and $B$ is a linear transformation. We now assign the data fidelity term $\frac{1}{2}\|\widetilde{A}f-\widetilde{m}\|_2^2$ as $L_1$,  the total variation regularization term $\alpha \|D f\|_{2, 1}$ as $(f_2 \circ B)$, and $f_3$ is the indicator function
\begin{equation}
    \label{eq:indicator_function}
    \chi_{\mathbb{R}_+^N} = 
    \begin{cases} 
        0, & f \in \mathbb{R}_+^N, \\  
        \infty, & f \notin \mathbb{R}_+^N. 
    \end{cases} 
\end{equation}

Adapting the PDFP algorithm to solve (\ref{eq:regularized_ct_tv_model}) yields the following iterations:
\begin{equation}
    \label{eq:pdfp_iterations}
    \begin{cases} 
        g^{\nu+1} 	&= \mathrm{proj}_{\mathbb{R}_+^N}(f^\nu  -  \gamma \widetilde{A}^T(\widetilde{A}f^\nu - \widetilde{m}) -  \lambda D^T v^\nu), \\
        v^{\nu+1} 	&= (I -\mathrm{prox}_{\frac{\gamma}{\lambda}\alpha\|\cdot\|_{2,1}})(D g^{\nu+1} + v^\nu), \\
        f^{\nu+1}	&= \mathrm{proj}_{\mathbb{R}_+^N}(f^\nu  -  \gamma \widetilde{A}^T(\widetilde{A}f^\nu - \widetilde{m})  -  \lambda D^T v^{\nu+1}),
    \end{cases}
\end{equation}
where $\nu$ is the iteration number, $0 < \lambda < 1/\lambda_\mathrm{max} (DD^T)$, $0 < \gamma < 2\beta$, $\mathrm{proj}_{\mathbb{R}_+^N}$ is the orthogonal projection onto $\mathbb{R}^N_+$, and
\begin{equation}
    \label{eq:l21_proximal_operator}
    \mathrm{prox}_{\frac{\gamma}{\lambda}\alpha\|\cdot\|_{2,1}} = \frac{Df}{\|Df\|_ 2}\max\{\|Df\|_2 - \frac{\gamma}{\lambda}\alpha , \,0\}
\end{equation}
is the proximal operator of the $\ell^{2, 1}$ mixed norm \cite{chen_et_al_2013__inverseprobl}, where the operations should be understood component-wise, keeping track of how the gradient components in $Df$ and the gradient magnitudes in $\|Df\|_2$ are associated with each other, and where the convention $\frac{0}{0} \cdot 0 = 0$ is used \cite{cil_optimization}. 

It is known that $\lambda_\mathrm{max} (DD^T) \leq 12$, which can be shown using the Gershgorin circle theorem \cite{esser_et_al_2010__siamjimagingsci}. The Lipschitz constant $K$ of $\nabla \frac{1}{2}\|\widetilde{A}f-\widetilde{m}\|_2^2 = \widetilde{A}^T(\widetilde{A}x-\widetilde{m})$ is 1, which follows directly from $\|\widetilde{A}\|_2 = 1$, resulting in $\beta = 1$. If the conditions for $\lambda$ and $\gamma$ are met, the iterations in (\ref{eq:pdfp_iterations}) will converge towards the minimum of (\ref{eq:regularized_ct_tv_model}).

\subsection{Controlled Gradient Sparsity}

Let $\kappa > 0$ and $f \in \mathbb{R}^N$. We define
\begin{equation}
    \label{eq:counting_function}
    \#_\kappa f = \#\{ l \; | \; 1 \leq l \leq N, \: |f_l| > \kappa \}.
\end{equation}
Here $\#_\kappa f$ counts the number of non-zero entries in $f$ using tolerance $\kappa$ around zero. This deals with extremely low-level numerical noise and floating-point arithmetic issues. The sparsity level of $f$ is defined as 
\begin{equation}
    \mathcal{C}_f = \frac{\#_\kappa f}{N},    
\end{equation}
\emph{i.e.} as the ratio of non-zero elements to all elements. Furthermore, the gradient sparsity level of $f$ is defined as
\begin{equation}
    \label{eq:gradient_sparsity}
    \mathcal{C}_{\nabla f} = \frac{\#_\kappa \| \nabla f \|_ 2}{N},
\end{equation}
using the definition of $\| \nabla f \|_ 2$ given Section \ref{sec:tv}. Eq. (\ref{eq:gradient_sparsity}) tracks the relative number of elements/voxels in $f$ with non-zero gradients.

In controlled gradient sparsity, we choose an \emph{a priori} sparsity level $\mathcal{C}_\mathrm{pr} \in (0, 1)$ and let $\alpha = \alpha^\nu$ vary during the iterations in (\ref{eq:pdfp_iterations}) so that the gradient sparsity level $C_{\nabla f}^\nu$ of the iterates $f^\nu$ converges to $\mathcal{C}_\mathrm{pr}$ as $\nu \rightarrow \infty$. Using ideas from control theory, the updating rule is set as
\begin{equation}
    \label{eq:alpha_update}
    \alpha^{\nu} = \max \{\alpha^{\nu-1} + \beta e^{\nu}, \, 0 \},
\end{equation}
where
\begin{equation}
    e^{\nu} = \mathcal{C}_{\nabla f}^{\nu-1} - \mathcal{C}_\mathrm{pr}
\end{equation}
and $\beta$ is called the \emph{tuning parameter}. This is based on the incremental PID-controller in control theory \cite{morari_1985__ieeetransautomcontrol}. The iterations are terminated if the relative distance between iterations 
$$
s^\nu = \frac{\|f^\nu - f^{\nu-1}\|_2}{\|f^\nu\|_2}
$$
falls below a preset threshold $s_\mathrm{min}$, or if a maximum number of iterations $\nu_\mathrm{max}$ is reached.

It can occur that the values of $\alpha^\nu$ become stuck at zero, in which case the reconstruction process becomes an unregularized problem equivalent to (\ref{eq:linear_measurement_model}). To avoid this, the iterations are interrupted if $\alpha^\nu$ reaches zero, and should be restarted with a smaller value of $\mathcal{C}_\mathrm{pr}$.

The resulting algorithm is called TV-CGS (\emph{Total Variation using Controlled Gradient Sparsity}). The pseudocode for TV-CGS is shown in Algorithm \ref{alg:tvcgs_v2}. We emphasize that unlike (\ref{eq:pdfp_iterations}), which is guaranteed to converge under sufficient conditions, there is not general proof of convergence for Algorithm \ref{alg:tvcgs_v2}, and we proceed to study its convergence properties heuristically via computational experiments.

\begin{algorithm}
    \caption{TV-CGS \\(\emph{Total Variation using Controlled Gradient Sparsity})}
    \label{alg:tvcgs_v2}
    \begin{algorithmic}
        \State \textbf{Input:} data $\widetilde{m}$, model $\widetilde{A}$, parameters $\mathcal{C}_\mathrm{pr}$, $\kappa$, $\beta$, $\alpha^0$, $\nu_\mathrm{max}$, $s_\mathrm{min}$, starting point $f^0$
        \State \textbf{Initialize: $\gamma \gets 1$, $\lambda \gets \frac{1}{13}$, $v^0 = Df^0$}, $\mathcal{C}_{\nabla f}^0 = 1$
        \For{$\nu \gets 1, \ldots, \nu_\mathrm{max}$}
            \State $e^{\nu} \gets \mathcal{C}_{\nabla f}^{\nu-1} - \mathcal{C}_\mathrm{pr}$
            \State $\alpha^\nu \gets \max \{\alpha^{\nu-1} + \beta e^\nu, \, 0 \}$
            \If{$\alpha^{\nu} = 0$}
                \State \textbf{interrupt}
            \EndIf
            \State $g^{\nu} \gets \mathrm{proj}_{\mathbb{R}_+^n} (f^{\nu-1} - \gamma \widetilde{A}^T(\widetilde{A}f^{\nu-1}-\widetilde{m})  -  \lambda D^T v^{\nu-1})$
            \State $v^{\nu} \gets (I - \mathrm{prox}_{\frac{\gamma}{\lambda}\alpha\|\cdot\|_{2,1}})(Dg^{\nu} + v^{\nu-1})$
            \State $f^{\nu} \gets \mathrm{proj}_{\mathbb{R}_+^n} (f^{\nu-1}  -  \gamma \widetilde{A}^T(\widetilde{A}f^{\nu-1}-\widetilde{m})  -  \lambda D^T v^{\nu})$
            \State $s^{\nu} \gets \frac{\|f^{\nu} - f^{\nu-1}\|_2}{\|f^{\nu}\|_2}$ 
            \State $\mathcal{C}_{\nabla f}^\nu \gets \#_\kappa \| \nabla f^\nu \|_ 2 / N$
            \If{$s^{\nu}< s_\mathrm{min}$}
                \State \textbf{break}
            \EndIf            
        \EndFor
        \State \Return $f^{\nu}$
    \end{algorithmic}
\end{algorithm}

\subsection{Dose and Noise Considerations}

The dominant source of noise in X-ray scans is the Poisson-distributed quantum noise from X-ray photon measurements. Other possible noise sources include detector noise arising from scintillators, A/D conversion, and electronic noise \cite{swank_1973__japplphys, buzug_2008}. 
The signal-to-noise ratio (SNR) for Poisson-distributed photon measurements is proportional to the square root of the number of photons used. Because dose is directly proportional to the number of X-ray photons, we have $SNR \propto \sqrt{\mathrm{dose}}$. Therefore, in clinical scanners it is possible to reduce the patient dose by reducing the tube current - time product (milli\-ampere-second, or mAs), at the expense of increased noise. TV-based algorithms are inherently denoising, with potential for dose reduction without compromising image quality.

\subsection{Computational Implementation}

All algorithms were implemented with the MATLAB programming language, and reconstructions were computed using MATLAB R2023b \cite{matlab_r2023b} on a Lenovo ThinkPad P16 Gen 2 PC workstation equipped with 32 GB of RAM and a 16 GB NVIDIA GeForce RTX 4090 graphics processing unit (GPU). All forward projections, backprojections, and FDK reconstructions were computed using the GPU. The CBCT imaging system model and FDK reconstructions were modelled and implemented using the ASTRA Toolbox v. 2.1.0 \cite{van_aarle_et_al_2016__optexpress, van_aarle_et_al_2015__ultramicroscopy}. The forward model $A$ was implemented as a GPU-based matrix-free linear operator using ASTRA and the Spot linear-operator toolbox \cite{van_den_berg_friedlander_2013__spot}. 

Throughout this work, we set $\lambda = 1/13$, $\gamma = 1$, $\kappa = 10^{-6}$, and $s_\mathrm{min} = 10^{-6}$. If the tuning parameter $\beta$ is too large, it can lead to oscillating solutions, whereas if it is sufficiently small, it will only affect the convergence speed of the algorithm. We heuristically set the value $\beta = 3 \cdot 10^{-7}$ throughout this work. The initial value of the regularization parameter should not in principle affect the end result of the algorithm, but it can affect the convergence speed. Throughout this work we set $\alpha^0 = 10^{-6}$, as we have discovered from subjective experience that this is the approximate order of magnitude of $\alpha$ for very mildly regularized solutions when using (\ref{eq:pdfp_iterations}) for CBCT reconstruction.

The values of $\mathcal{C}_\mathrm{pr}$ and $\nu_\mathrm{max}$ were set depending on the experiment.

\subsection{Experiments Using Simulated Data}

The convergence behaviour of the TV-CGS algorithm was studied using the 3D Shepp-Logan phantom, as it usefully provides an absolute ground truth for reference \cite{schabel_shepplogan3d}. We implemented a $256 \times 256 \times 256$ phantom, modelled the voxel size as \SI{0.75}{mm}, and scaled the voxel grayscale values so that the maximum linear attenuation coefficient in the phantom is equivalent to that of cortical bone at \SI{60}{keV} ($\mu = 0.0453312 \;1/\mathrm{voxel}$)   \cite{nist_database_126}. All X-ray measurements were simulated with \SI{60}{keV} monochromatic radiation.

We modelled a CBCT system with a $256 \times 256$ flat panel detector with a pixel size of \SI{1.20}{mm}, a source-to-center-of-rotation distance of \SI{500}{mm}, and a source-to-detector distance of \SI{800}{mm}. 
For each scan, we simulated 900 evenly spaced projections on the interval $[0\degree, \, 359.6\degree ]$. In order to avoid committing inverse crime when simulating the data, \emph{i.e.}, using the same forward model and discretization for the data generation and reconstruction \cite{kaipio_somersalo_2005, kaipio_somersalo_2007__jcomputapplmath}, we rotated the phantom by the angle $\theta_0 = \frac{e}{\pi^2} \cdot 180\degree$, and offset each projection direction with $\theta_0 + \delta\theta$, where $\delta\theta$ are random perturbations chosen from the uniform continuous distribution $[-0.01\degree, \, 0.01\degree]$.

Photon noise with a Poisson distribution was simulated in the projection data by determining the expectation values for the photon counts at the detector pixels using the formula
\begin{equation}
    E(I) = \frac{I_0 r_0^2}{r^2} e^{-Af},
\end{equation}
where $I_0$ is a parameter defining the number of expected counts on a pixel-sized area at distance $r_0$ from the source, $r$ keeps track of the source-to-detector-pixel distances, $A$ is the inverse crime -free forward model, and $f$ is the rotated phantom. Other noise sources or scattering were not modelled. A single flat-field image was created by simulating and averaging 400 X-ray projections without the phantom, and the data was log-transformed into standard form as defined in Eq. [\ref{eq:lambert_beer}].

We simulated a set of CBCT scans using $I_0 \in \{250, 500, 1000, 1500, 2000, 4000, 5000\}$. We computed reconstructions using FDK and TV-CGS with $\mathcal{C}\mathrm{pr} \in \{$0.05, 0.075, 0.10, 0.125, 0.15, 0.175, 0.2, 0.225, 0.25, 0.275, 0.30\} and $\nu_\mathrm{max} = 5000$. We tracked the convergence of $s^\nu$, $\mathcal{C}_{\nabla f}^\nu$, $\alpha^\nu$, and the $\ell^2$ distance between the iterations and the ground truth.

\section{Results}

\subsection{Simulated data}

\begin{figure}[ht]
    \centering
    \includegraphics[width=0.7\columnwidth]{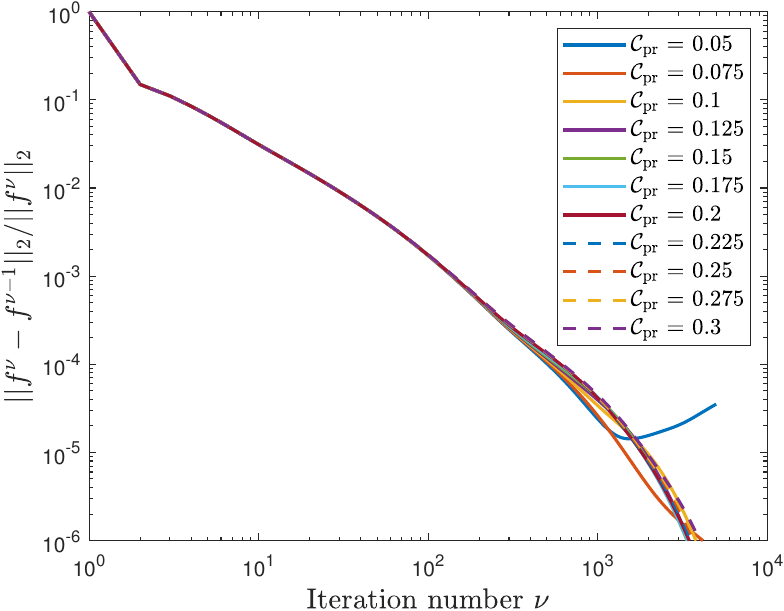}
    \caption{Convergence of the relative distance between consecutive TV-CGS iterations for the 3D Shepp-Logan phantom when $N_1 = N_2 = N_3 = 256$ and $I_0 = 1000$.}
    \label{fig:shepp_logan_3d_iteration_distances_constant_dose}
\end{figure}

Figure \ref{fig:shepp_logan_3d_iteration_distances_constant_dose} shows the convergence of the relative distance between consecutive iterations $s^\nu = \|f^\nu - f^{\nu - 1}\|_2 / \|f^\nu\|_2$ at a constant dose level ($I_0 = 1000$) and varying prior sparsity levels $\mathcal{C}_\mathrm{pr}$. For $\mathcal{C}_\mathrm{pr} = 0.05$ the TV-CGS algorithm is unstable and begins to diverge after approximately 1000 iterations. For all other values of $\mathcal{C}_\mathrm{pr}$ the values of $s^\nu$ are monotonously decreasing, and the minimum tolerance level $s_\mathrm{min} = 10^{-6}$ was reached before the maximum number of iterations $\nu_\mathrm{max}$. 

\begin{figure}[ht]
    \centering

    \begin{subfigure}[b]{0.7\columnwidth}
        \centering
        \includegraphics[width=\columnwidth]{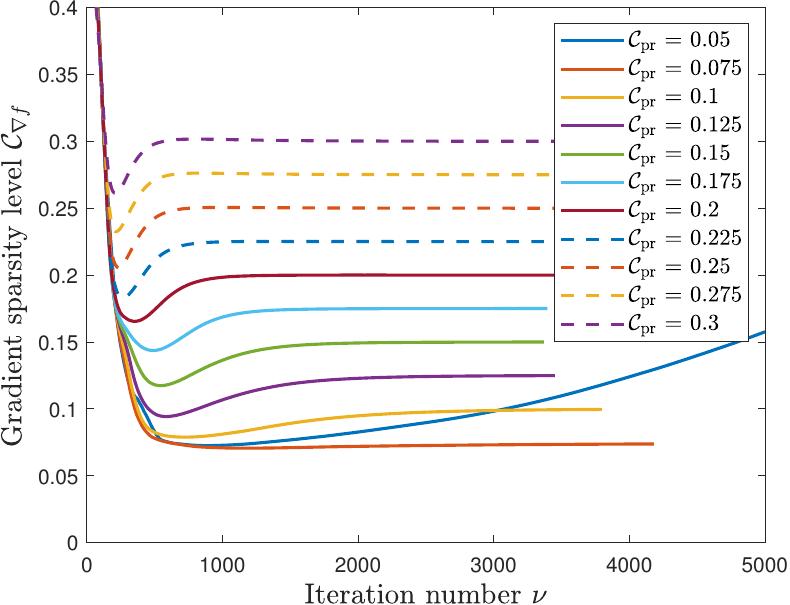}
        \caption{}
        \label{fig:shepp_logan_3d_gradient_sparsities_constant_dose}
    \end{subfigure}

    \begin{subfigure}[b]{0.7\columnwidth}
        \centering
        \includegraphics[width=\columnwidth]{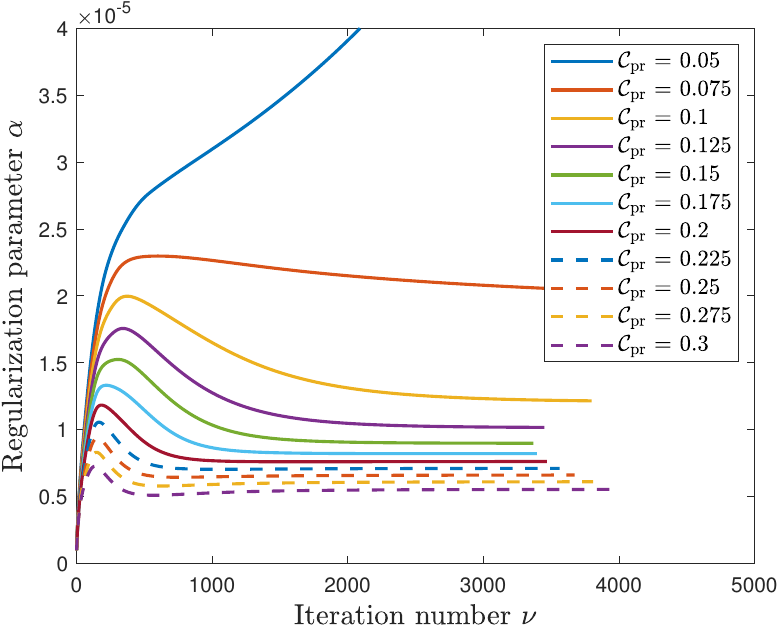}
        \caption{}
        \label{fig:shepp_logan_3d_regularization_parameters_constant_dose}
    \end{subfigure}
    
    \caption{Convergence of a) $\mathcal{C}_{\nabla f}$, and b) $\alpha$ for the 3D Shepp-Logan phantom when $N_1 = N_2 = N_3 = 256$ and $I_0 = 1000$.}
    \label{fig:shepp_logan_3d_sparsities_parameters_constant_dose}
\end{figure}

Figure \ref{fig:shepp_logan_3d_sparsities_parameters_constant_dose} shows the convergence of the gradient sparsity level $\mathcal{C}_\mathrm{\nabla f}^\nu$ and the regularization parameter $\alpha^\nu$ at the constant dose level and varying prior sparsity levels $\mathcal{C}_\mathrm{pr}$. In all cases except for $\mathcal{C}_\mathrm{pr} = 0.05$, both values converge to a constant level well before the iteration procedure was terminated.

\begin{figure}[ht]
    \centering
    \begin{overpic}[width=0.95\columnwidth]{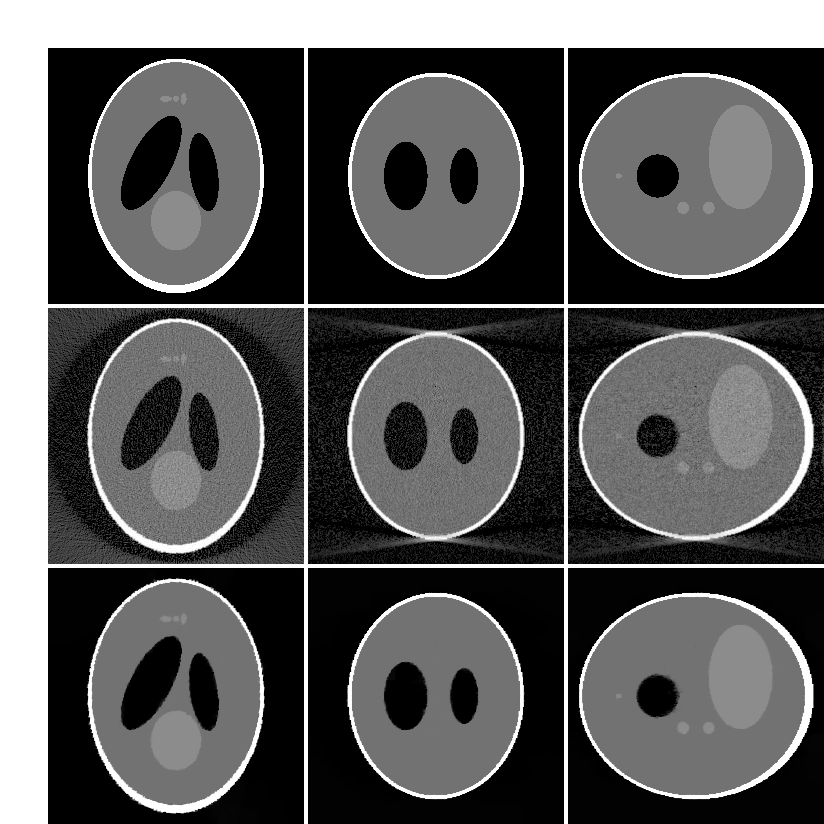}
        \put(16, 96){Axial}
        \put(46, 96){Coronal}
        \put(78, 96){Sagittal}
        \put(1.4,68){\rotatebox{90}{Ground Truth}}
        \put(1.4,43.5){\rotatebox{90}{FDK}}
        \put(1.4,8.5){\rotatebox{90}{TV-CGS}}
    \end{overpic}
    \caption{Center slices in the axial, coronal, and sagittal directions of the ground truth, the FDK reconstruction, and the TV-CGS reconstruction with $I_0 = 1000$ and $\mathcal{C}_\mathrm{pr} = 0.15$.}
    \label{fig:shepp_logan_3d_reconstruction_collage}
\end{figure}

Figure \ref{fig:shepp_logan_3d_reconstruction_collage} shows the ground truth, the FDK reconstruction for $I_0 = 1000$, and the TV-CGS reconstruction for $I_0 = 1000$ and $\mathcal{C}_\mathrm{pr} = 0.15$, which was the fastest to converge to tolerance. The FDK reconstruction contains significant noise and cone beam artifacts, whereas the TV-CGS reconstruction is very close to the ground truth, with negligible artifacts, very little noise, and increased visibility of small features in the phantom.

\begin{figure}[ht]
    \centering

    \begin{subfigure}[b]{0.7\columnwidth}
        \centering
        \includegraphics[width=\columnwidth]{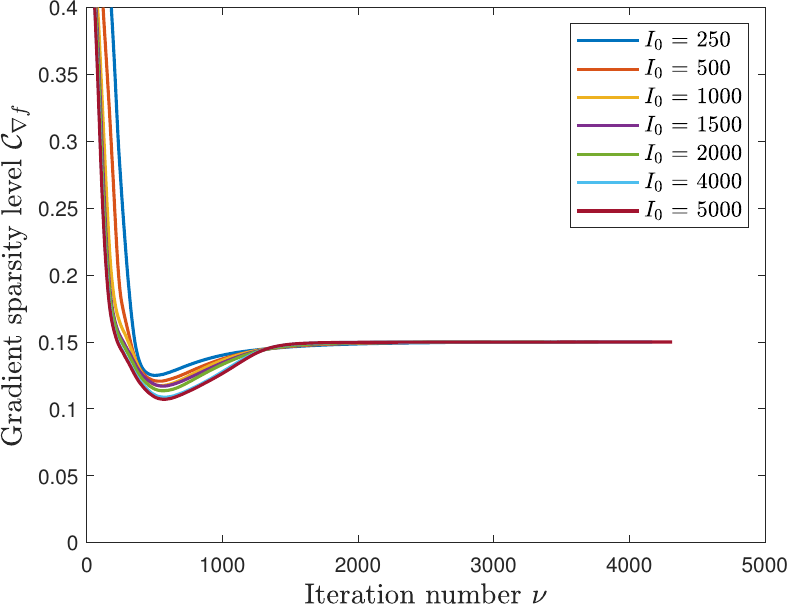}
        \caption{}
        \label{fig:shepp_logan_3d_gradient_sparsities_constant_sparsity}
    \end{subfigure}

    \begin{subfigure}[b]{0.7\columnwidth}
        \centering
        \includegraphics[width=\columnwidth]{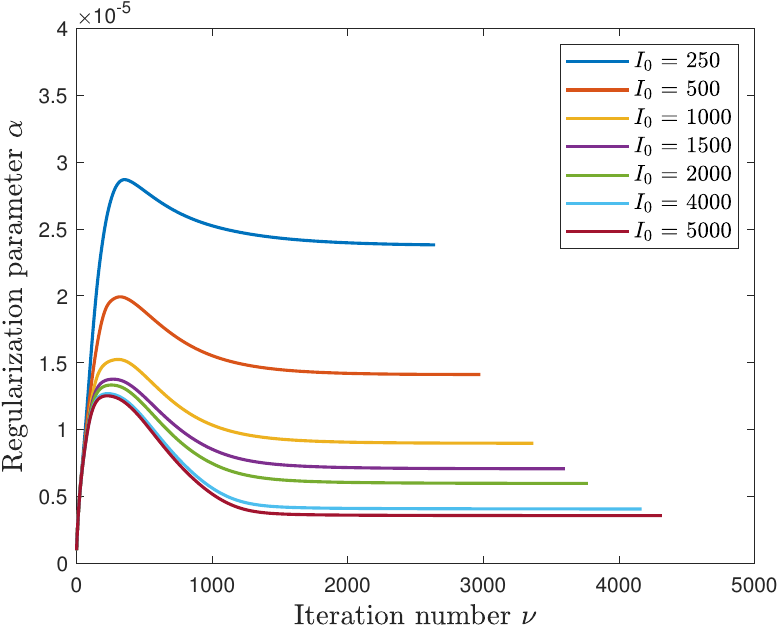}
        \caption{}
        \label{fig:shepp_logan_3d_regularization_parameters_constant_sparsity}
    \end{subfigure}
    
    \caption{Convergence of a) $\mathcal{C}_{\nabla f}$, and b) $\alpha$ for the 3D Shepp-Logan phantom when $N_1 = N_2 = N_3 = 256$ and $\mathcal{C}_\mathrm{pr} = 0.15$.}
    \label{fig:shepp_logan_3d_sparsities_parameters_constant_sparsity}
\end{figure}

Figure \ref{fig:shepp_logan_3d_sparsities_parameters_constant_sparsity} shows the convergence of $\mathcal{C}_\mathrm{\nabla f}^\nu$ and $\alpha^\nu$ with constant sparsity ($\mathcal{C}_\mathrm{pr} = 0.15$) and varying dose levels. In all cases except $\mathcal{C}_\mathrm{pr} = 0.05$, TV-CGS converges to the specified gradient sparsity level and to constant value of the regularization parameter which is increased with higher noise levels.

We note that for the (unrotated) ground truth Shepp-Logan 3D phantom at this resolution the true gradient sparsity is $0.0197$. It would appear that stable reconstruction is achieved at a slightly higher sparsity level, which is not surprising considering that modelling error was included in the simulations.

The average iteration duration in seconds was tracked for all reconstructions. For the computations where the dose level was kept constant and the gradient sparsity was varied, the average of the iterations duration averages was $4.34 \pm 0.10$ seconds. For the computations where the sparsity was constant and the dose level was varied, the average of the iterations duration averages was also $4.18 \pm 0.24$ seconds.

\section{Discussion}

We introduce a total variation reconstruction method for CBCT with automatic parameter choice. The approach is based on changing the regularization parameter dynamically during PDFP iteration so that the resulting reconstruction has predescribed sparsity. The adaptive adjustment of parameter on the fly is based on basic control theory. Controlled sparsity of the image gradient magnitude offers increased interpretability of the regularization strength: the parameter $\mathcal{C}_\mathrm{pr}$ has an intuitive interpretation, and may be easier to adjust as needed compared to $\alpha$, whose numerical value carries little immediate interpretability in itself.

We remark that there is no convergence proof for the full nonlinear algorithm. However, with a fixed regularization parameter we know that the PDFP theory of \cite{chen_et_al_2016__jfixpointtheorya} guarantees convergence. We do observe in almost all numerical tests that the regularization parameter settles to an asymptotic value after initial dynamic adjustments, and the unstable case was an extreme outlier. One could therefore freeze the adjustment of the parameter after enough iterations and then enter theory-covered behavior regime. 

A possible criticism towards our approach is the fact that while the value of $\alpha$ is automatically adjusted, we still need to choose the tuning parameter $\beta$. However, the choice of $\beta$ is immensely simpler than the notoriously vicious choice of $\alpha$. Namely, too large $\beta$ causes easily detectable divergence in the iteration, while too small $\beta$ makes the iteration converge slower. It is rather straightworward to pick a good $\beta$ in a given application.

\clearpage
\bibliographystyle{unsrt}
\bibliography{tvcgs_bibliography}

\begin{thebibliography}{10}

\bibitem{fahrig_et_al_2021__jmedimaging}
R.~{Fahrig}, D.~A. {Jaffray}, I.~{Sechopoulos}, and J.~W. {Stayman}.
\newblock Flat-panel conebeam {CT} in the clinic: history and current state.
\newblock {\em Journal of Medical Imaging}, 8(5):052115, 2021.

\bibitem{kaasalainen_et_al_2021__physmed}
T.~{Kaasalainen}, M.~{Ekholm}, T.~{Siiskonen}, and M.~{Kortesniemi}.
\newblock Dental cone beam {CT}: An updated review.
\newblock {\em Physica Medica}, 88:193--217, 2021.

\bibitem{feldkamp_et_al_1984__joptsocamera}
L.~A. {Feldkamp}, L.~C. {Davis}, and J.~W. {Kress}.
\newblock Practical cone-beam algorithm.
\newblock {\em Journal of the Optical Society of America A}, 1(6):612--619,
  1984.

\bibitem{gardner_et_al_2019__advradiatoncol}
S.~J. {Gardner}, W.~{Mao}, C.~{Liu}, I.~{Aref}, M.~{Elshaikh}, J.~K. {Lee},
  D.~{Pradhan}, B.~{Movsas}, I.~J. {Chetty}, and F.~{Siddiqui}.
\newblock Improvements in {CBCT} image quality using a novel iterative
  reconstruction algorithm: A clinical evaluation.
\newblock {\em Advances in Radiation Oncology}, 4:390--400, 2019.

\bibitem{ravishankar_et_al_2020__procieee}
S.~{Ravishankar}, J.~C. {Ye}, and J.~A. {Fessler}.
\newblock Image reconstruction: From sparsity to data-adaptive methods and
  machine learning.
\newblock {\em Proceedings of the IEEE}, 108(1):86--–109, 2020.

\bibitem{matenine_et_al_2020__neuroradiology}
D.~{Matenine}, M.~{Kachelriess}, P.~{Despr\'es}, J.~A. {de Guise}, and
  M.~{Schmittbuhl}.
\newblock Potential of iterative reconstruction for maxillofacial cone beam
  {CT} imaging: technical note.
\newblock {\em Neuroradiology}, 62:1511--1514, 2020.

\bibitem{wu_et_al_2020__medphys}
P.~{Wu}, A.~{Sisniega}, J.~W. {Stayman}, W.~{Zbijewski}, D.~{Foos}, X.~{Wang},
  N.~{Khanna}, N.~{Aygun}, R.~D. {Stevens}, and J.~H. {Siewerdsen}.
\newblock Cone-beam {CT} for imaging of the head/brain: Development and
  assessment of scanner prototype and reconstruction algorithms.
\newblock {\em Medical Physics}, 47(6):2392--2407, 2020.

\bibitem{lagerwerf_et_al_2020__jimaging}
M.~J. {Lagerwerf}, D.~M. {Pelt}, W.~J. {Palenstijn}, and K.~J. {Batenburg}.
\newblock A computationally efficient reconstruction algorithm for circular
  cone-beam computed tomography using shallow neural networks.
\newblock {\em Journal of Imaging}, 6(12):135, 2020.

\bibitem{lu_et_al_2021__physmedbiol}
K.~{Lu}, L.~{Ren}, and F.-F. {Yin}.
\newblock A geometry-guided deep learning technique for {CBCT} reconstruction.
\newblock {\em Physics in Medicine \& Biology}, 66(15), 2021.

\bibitem{moriakov_et_al_2023__medphys}
N.~{Moriakov}, J.-J. {Sonke}, and J.~{Teuwen}.
\newblock End-to-end memory-efficient reconstruction for cone beam {CT}.
\newblock {\em Medical Physics}, 50:7579--7593, 2023.

\bibitem{rudin_et_al_1992__physicad}
L.~I. {Rudin}, S.~{Osher}, and E.~{Fatemi}.
\newblock Nonlinear total variation based noise removal algorithms.
\newblock {\em Physica D}, 60(1--4):259--268, 1992.

\bibitem{delaney_bresler_1998__ieeetransimageprocess}
A.~H. {Delaney} and Y.~{Bresler}.
\newblock Globally convergent edge-preserving regularized reconstruction: an
  application to limited-angle tomography.
\newblock {\em IEEE Transactions on Image Processing}, 7(2):204--221, 1998.

\bibitem{persson_et_al_2001__physmedbiol}
M.~{Persson}, D.~{Bone}, and H.~{Elmqvist}.
\newblock Total variation norm for three-dimensional iterative reconstruction
  in limited view angle tomography.
\newblock {\em Physics in Medicine \& Biology}, 46(3):853--866, 2001.

\bibitem{kolehmainen_et_al_2003__physmedbiol}
V.~{Kolehmainen}, S.~{Siltanen}, S.~{J\"arvenp\"a\"a}, J.~P. {Kaipio},
  P.~{Koistinen}, M.~{Lassas}, J.~{Pirttil\"a}, and E.~{Somersalo}.
\newblock Statistical inversion for medical x-ray tomography with few
  radiographs: {II}. application to dental radiology.
\newblock {\em Physics in Medicine \& Biology}, 48(10):1465--1490, 2003.

\bibitem{kolehmainen_et_al_2006__ieeetransmedimaging}
V.~{Kolehmainen}, A.~{Vanne}, S.~{Siltanen}, S.~J{\"a}rvenp{\"a}{\"a}, J.~P.
  {Kaipio}, M.~{Lassas}, and M.~{Kalke}.
\newblock Parallelized {Bayesian} inversion for three-dimensional dental
  {X-ray} imaging.
\newblock {\em IEEE Transactions on Medical Imaging}, 25(2):218--228, 2006.

\bibitem{sidky_et_al_2006__nssmic}
E.~Y. {Sidky}, C.~M. {Kao}, and {X.} Pan.
\newblock Effect of the data constraint on few-view, fan-beam {CT} image
  reconstruction by {TV} minimization.
\newblock In {\em 2006 IEEE Nuclear Science Symposium Conference Record},
  volume~4, pages 2296--2298, 2006.

\bibitem{sidky_et_al_2006__jxrayscitechnol}
E.~Y. {Sidky}, C.~M. {Kao}, and X.~{Pan}.
\newblock Accurate image reconstruction from few-views and limited-angle data
  in divergent-beam {CT}.
\newblock {\em Journal of {X}-Ray Science and Technology}, 14(2):119--139,
  2006.

\bibitem{liao_sapiro_2008__isbi}
H.~Y. {Liao} and G.~{Sapiro}.
\newblock Sparse representations for limited data tomography.
\newblock In {\em 5th IEEE International Symposium on Biomedical Imaging: From
  Nano to Macro}, pages 1375--1378, 2008.

\bibitem{herman_davidi_2008__inverseprobl}
G.~T. {Herman} and R.~{Davidi}.
\newblock Image reconstruction from a small number of projections.
\newblock {\em Inverse Problems}, 24(4):045011, 2008.

\bibitem{tang_et_al_2009__physmedbiol}
J.~{Tang}, B.~E. {Nett}, and G.~H. {Chen}.
\newblock Performance comparison between total variation ({TV})-based
  compressed sensing and statistical iterative reconstruction algorithms.
\newblock {\em Physics in Medicine \& Biology}, 54(19):5781--5804, 2009.

\bibitem{duan_et_al_2009__ieeetransnuclsci}
X.~{Duan}, L.~{Zhang}, Y.~{Xing}, Z.~{Chen}, and J.~{Cheng}.
\newblock Few-view projection reconstruction with an iterative
  reconstruction-reprojection algorithm and {TV} constraint.
\newblock {\em IEEE Transactions on Nuclear Science}, 56(3):1377--1382, 2009.

\bibitem{bian_et_al_2010__tsinghuascitechnol}
J.~{Bian}, X.~{Han}, E.~Y. {Sidky}, G.~{Cao}, J.~{Lu}, O.~{Zhou}, and X.~{Pan}.
\newblock Investigation of sparse data mouse imaging using micro-{CT} with a
  carbon-nanotube-based {X}-ray source.
\newblock {\em Tsinghua Science and Technology}, 15(1):74--78, 2010.

\bibitem{bian_et_al_2010__physmedbiol}
J.~{Bian}, J.~H. {Siewerdsen}, X.~{Han}, E.~Y. {Sidky}, J.~L. {Prince}, C.~A.
  {Pelizzari}, and X.~{Pan}.
\newblock Evaluation of sparse-view reconstruction from flat-panel-detector
  cone-beam {CT}.
\newblock {\em Physics in Medicine \& Biology}, 55(22):6575--6599, 2010.

\bibitem{tian_et_al_2011__physmedbiol}
Z.~{Tian}, X.~{Jia}, K.~{Yuan}, T.~{Pan}, and S.~B. {Jiang}.
\newblock Low-dose {CT} reconstruction via edge-preserving total variation
  regularization.
\newblock {\em Physics in Medicine \& Biology}, 56(18):5949--5967, 2011.

\bibitem{jensen_et_al_2012__bitnumermath}
T.~L. {Jensen}, J.~H. {J{\o}rgensen}, P.~C. {Hansen}, and S.~H. {Jensen}.
\newblock Implementation of an optimal first-order method for strongly convex
  total variation regularization.
\newblock {\em BIT Numerical Mathematics}, 52:329--356, 2012.

\bibitem{hamalainen_et_al_2014__intjtomogrsimul}
K.~{H\"am\"al\"ainen}, L.~{Harhanen}, A.~{Hauptmann}, A.~{Kallonen},
  E.~{Niemi}, and S.~{Siltanen}.
\newblock Total variation regularization for large-scale {X}-ray tomography.
\newblock {\em International Journal of Tomography and Simulation}, 25(1),
  2014.

\bibitem{jorgensen_sidky_2015__philostransrsoca}
J.~S. {J{\o}rgensen} and E.~Y. {Sidky}.
\newblock How little data is enough? {Phase-diagram} analysis of
  sparsity-regularized {X-ray} computed tomography.
\newblock {\em Philosophical Transactions of the Royal Society A},
  373:20140387, 2015.

\bibitem{yang_et_al_2017__physmedbiol}
C.~{Yang}, P.~{Wu}, S.~{Gong}, J.~{Wang}, Q.~{Lyu}, X.~{Tang}, and T.~{Niu}.
\newblock Shading correction assisted iterative cone-beam {CT} reconstruction.
\newblock {\em Physics in Medicine \& Biology}, 62(22):8495--8520, 2017.

\bibitem{chen_et_al_2018__physmedbiol}
Y.~{Chen}, F.~F. {Yin}, Y.~{Zhang}, Y.~{Zhang}, and L.~{Ren}.
\newblock Low dose {CBCT} reconstruction via prior contour based total
  variation ({PCTV}) regularization: a feasibility study.
\newblock {\em Physics in Medicine \& Biology}, 63(8):085014, 2018.

\bibitem{tseng_et_al_2020__physmed}
H.~W. {Tseng}, S.~{Vedanthama}, and A.~{Karellas}.
\newblock Cone-beam breast computed tomography using ultra-fast image
  reconstruction with onstrained, total-variation minimization for suppression
  of artifacts.
\newblock {\em Physica Medica}, 73:117--124, 2020.

\bibitem{tseng_et_al_2022__physmedbiol}
H.~W. {Tseng}, A.~{Karellas}, and S.~{Vedanthama}.
\newblock Cone-beam breast {CT} using an offset detector: effect of detector
  offset and image reconstruction algorithm.
\newblock {\em Physics in Medicine \& Biology}, 67(8), 2022.

\bibitem{sabate_landman_2023__physmedbiol}
M.~{Sabat{\'e} Landman}, A.~{Biguri}, S.~{Hatamikia}, R.~{Boardman},
  J.~{Aston}, and C.-B. {Sch{\"o}nlieb}.
\newblock On {Krylov} methods for large-scale {CBCT} reconstruction.
\newblock {\em Physics in Medicine \& Biology}, 68(15):155008, 2023.

\bibitem{zhang_et_al_2023__medphys}
P.~{Zhang}, S.~{Ren}, Y.~{Liu}, Z.~{Gui}, H.~{Shangguan}, Y.~{Wang}, H.~{Shu},
  and Y.~{Chen}.
\newblock A total variation prior unrolling approach for computed tomography
  reconstruction.
\newblock {\em Medical Physics}, 50(5):2816--2834, 2023.

\bibitem{clason_et_al_2010__siamjscicomput}
C.~{Clason}, B.~{Jin}, and K.~{Kunisch}.
\newblock A duality-based splitting method for $\ell^1$-{TV} image restoration
  with automatic regularization parameter choice.
\newblock {\em SIAM Journal on Scientific Computing}, 32(3):1484--1505, 2010.

\bibitem{dong_et_al_2011__jmathimagingvis}
Y.~{Dong}, M.~{Hinterm\"uller}, and M.~M. {Rincon-Camacho}.
\newblock Automated regularization parameter selection in multi-scale total
  variation models for image restoration.
\newblock {\em Journal of Mathematical Imaging and Vision}, 40:82--104, 2011.

\bibitem{wen_chan_2012__ieeetransimageprocess}
Y.~W. {Wen} and R.~H. {Chan}.
\newblock Parameter selection for total-variation-based image restoration using
  discrepancy principle.
\newblock {\em IEEE Transactions on Image Processing}, 21(4):1770--1781, 2012.

\bibitem{kolehmainen_et_al_2012__inverseprobl}
V.~{Kolehmainen}, M.~{Lassas}, K.~{Niinim{\"a}ki}, and S.~{Siltanen}.
\newblock Sparsity-promoting {Bayesian} inversion.
\newblock {\em Inverse Problems}, 28(2):025005, 2012.

\bibitem{frick_et_al_2012__electronjstat}
K.~{Frick}, P.~{Marnitz}, and A.~{Munk}.
\newblock Statistical multiresolution {Dantzig} estimation in imaging:
  Fundamental concepts and algorithmic framework.
\newblock {\em Electronic Journal of Statistics}, 6:231--268, 2012.

\bibitem{kindermann_et_al_2014__jinverseillposedprobl}
S.~{Kindermann}, L.~D. {Mutimbu}, and E.~{Resmerita}.
\newblock A numerical study of heuristic parameter choice rules for total
  variation regularization.
\newblock {\em Journal of Inverse and Ill-Posed Problems}, 22(1):63--94, 2014.

\bibitem{chen_et_al_2014__numeralgor}
K.~{Chen}, E.~{Loli Piccolomini}, and F.~{Zama}.
\newblock An automatic regularization parameter selection algorithm in the
  total variation model for image deblurring.
\newblock {\em Numerical Algorithms}, 67:73--92, 2014.

\bibitem{toma_et_al_2015__inverseproblimaging}
A.~{Toma}, B.~{Sixou}, and F.~{Peyrin}.
\newblock Iterative choice of the optimal regularization parameter in {TV}
  image restoration.
\newblock {\em Inverse Problems and Imaging}, 9(4):1171--1191, 2015.

\bibitem{niinimaki_et_al_2016__siamjimagingsci}
K.~{Niinim\"aki}, M.~{Lassas}, K.~{H\"am\"al\"ainen}, A.~{Kallonen},
  V.~{Kolehmainen}, E.~{Niemi}, and S.~{Siltanen}.
\newblock Multiresolution parameter choice method for total variation
  regularized tomography.
\newblock {\em SIAM Journal of Imaging Sciences}, 9(3):938--974, 2016.

\bibitem{hamalainen_et_al_2013__siamjscicomput}
K.~{H\"am\"al\"ainen}, A.~{Kallonen}, V.~{Kolehmainen}, M.~{Lassas},
  K.~{Niinim\"aki}, and S.~{Siltanen}.
\newblock Sparse tomography.
\newblock {\em SIAM Journal on Scientific Computing}, 35(3):B644--B665, 2013.

\bibitem{purisha_et_al_2018__measscitechnol}
Z.~{Purisha}, J.~{Rimpel{\"a}inen}, T.~{Bubba}, and S.~{Siltanen}.
\newblock Controlled wavelet domain sparsity for x-ray tomography.
\newblock {\em Measurement Science and Technology}, 29(1):014002, 2018.

\bibitem{chen_et_al_2016__jfixpointtheorya}
P.~{Chen}, J.~{Huang}, and X.~{Zhang}.
\newblock A primal-dual fixed point algorithm for minimization of the sum of
  three convex separable functions.
\newblock {\em Fixed Point Theory and Applications}, 2016:54, 2016.

\bibitem{kak_slaney_1988}
A.~C. {Kak} and M.~{Slaney}.
\newblock {\em Principles of Computerized Tomographic Imaging}.
\newblock IEEE Press, 1988.

\bibitem{buzug_2008}
T.~M. {Buzug}.
\newblock {\em Computed Tomography: From Photon Statistics to Modern Cone-Beam
  {CT}}.
\newblock Springer, 2008.

\bibitem{park_et_al_2015__philostransrsoca}
H.~S. {Park}, Y.~E. {Chung}, and J.~K. {Seo}.
\newblock Computed tomographic beam-hardening artefacts: mathematical
  characterization and analysis.
\newblock {\em Philosophical Transactions of the Royal Society A: Mathematical,
  Physical and Engineering Sciences}, 373(2043):20140388, 2015.

\bibitem{natterer_1986}
F.~{Natterer}.
\newblock {\em The Mathematics of Computerized Tomography}.
\newblock John Wiley \& Sons, 1986.

\bibitem{getreuer_2012__ipol}
P.~{Getreuer}.
\newblock {Rudin–Osher–Fatemi} total variation denoising using split
  {Bregman}.
\newblock {\em Image Processing On Line}, 2:74--95, 2012.

\bibitem{condat_2017__siamjimagingsci}
L.~{Condat}.
\newblock Discrete total variation: New definition and minimization.
\newblock {\em SIAM Journal on Imaging Sciences}, 10(3):1258--1290, 2017.

\bibitem{gramfort_et_al_2012__physmedbiol}
A.~{Gramfort}, M.~{Kowalski}, and M.~{H\"am\"al\"ainen}.
\newblock Mixed-norm estimates for the {M/EEG} inverse problem using
  accelerated gradient methods.
\newblock {\em Physics in Medicine \& Biology}, 57(7):1937--1961, 2012.

\bibitem{chen_et_al_2013__inverseprobl}
P.~{Chen}, J.~{Huang}, and X.~{Zhang}.
\newblock A primal–dual fixed point algorithm for convex separable
  minimization with applications to image restoration.
\newblock {\em Inverse Problems}, 29:025011, 2013.

\bibitem{cil_optimization}
{Core Imaging Library (CIL)} documentation: Optimisation framework.
\newblock \url{https://tomographicimaging.github.io/CIL/nightly/optimisation/}.
\newblock (Accessed: 2024-02-15).

\bibitem{esser_et_al_2010__siamjimagingsci}
E.~{Esser}, X.~{Zhang}, and T.~{Chan}.
\newblock A general framework for a class of first order primal-dual algorithms
  for convex optimization in imaging science.
\newblock {\em SIAM Journal on Imaging Sciences}, 3(4):1015--1046, 2010.

\bibitem{morari_1985__ieeetransautomcontrol}
M.~{Manfred}.
\newblock Robust stability of systems with integral control.
\newblock {\em IEEE Transactions on Automatic Control}, 30(6):574--577, 1985.

\bibitem{swank_1973__japplphys}
R.~K. {Swank}.
\newblock Absorption and noise in x‐ray phosphors.
\newblock {\em Journal of Applied Physics}, 44(9):4199--4203, 1973.

\bibitem{matlab_r2023b}
The~MathWorks Inc.
\newblock Matlab version: 23.2 (r2023b).
\newblock \url{https://www.mathworks.com}, 2023.

\bibitem{van_aarle_et_al_2016__optexpress}
W.~{van Aarle}, W.~J. {Palenstijn}, J.~{Cant}, E.~{Janssens}, F.~{Bleichrodt},
  A.~{Dabravolski}, J.~{De Beenhouwer}, K.~J. {Batenburg}, and J.~{Sijbers}.
\newblock Fast and flexible {X-ray} tomography using the {ASTRA} toolbox.
\newblock {\em Optics Express}, 24(22):25129--25147, 2016.

\bibitem{van_aarle_et_al_2015__ultramicroscopy}
W.~{van Aarle}, W.~J. {Palenstijn}, J.~{De Beenhouwer}, T.~{Altantzis},
  S.~{Bals}, K.~J. {Batenburg}, and J.~{Sijbers}.
\newblock The {ASTRA Toolbox}: A platform for advanced algorithm development in
  electron tomography.
\newblock {\em Ultramicroscopy}, 157:35--47, 2015.

\bibitem{van_den_berg_friedlander_2013__spot}
E.~{van den Berg} and M.~P. {Friedlander}.
\newblock Spot – a linear-operator toolbox, version 1.2.
\newblock \url{https://www.cs.ubc.ca/labs/scl/spot/index.html}, 2013.
\newblock (Accessed: 2024-02-09).

\bibitem{schabel_shepplogan3d}
Matthias {Schabel}.
\newblock {3D} {Shepp-Logan} phantom.
\newblock
  \url{https://www.mathworks.com/matlabcentral/fileexchange/9416-3d-shepp-logan-phantom}.
\newblock MATLAB Central File Exchange. (Accessed: 2024-04-26).

\bibitem{nist_database_126}
J.~H. {Hubbell} and S.~M. {Seltzer}.
\newblock X-ray mass attenuation coefficients - {NIST} standard reference
  database 126.
\newblock \url{https://www.nist.gov/pml/x-ray-mass-attenuation-coefficients},
  2004.
\newblock (Accessed: 2024-04-26).

\bibitem{kaipio_somersalo_2005}
J.~{Kaipio} and E.~{Somersalo}.
\newblock {\em Statistical and Computational Inverse Problems}.
\newblock Springer, 2005.

\bibitem{kaipio_somersalo_2007__jcomputapplmath}
J.~{Kaipio} and E.~{Somersalo}.
\newblock Statistical inverse problems: Discretization, model reduction and
  inverse crimes.
\newblock {\em Journal of Computational and Applied Mathematics}, 198:493--504,
  2007.

\end{thebibliography}

\end{document}